\documentclass[a4paper,11pt]{article}
\usepackage{pos}

\usepackage{slashed}
\usepackage{hyperref}
\usepackage{amsthm}
\usepackage{mathtools}
\usepackage{graphicx}

\usepackage{amssymb}

\usepackage{amsmath}
\usepackage{amsfonts}
\usepackage{simplewick}
\usepackage{tikz-cd}
\usetikzlibrary{decorations.markings}
\usetikzlibrary{intersections}
\usetikzlibrary{calc}
\usepackage{float}

\usepackage{bbm}

\def\swzero{{\textrm{\tiny $(0)$}}}

\def\1PI{{\textrm{\tiny 1PI}}}

\newcommand{\sfDelta}{{\rm{\Delta}}}
\newcommand{\sfLambda}{{\rm{\Lambda}}}

\newcommand{\CF}{\mathcal{F}}

\newcommand{\ii}{{\mathrm{i}}}

\newcommand{\sfR}{\mathrm{R}}

\newcommand{\nn}{\nonumber}
\def\d{{\rm d}}

\def\RR{{\mathcal R}}

\newcommand{\sfh}{{\sf h}}

\newcommand{\sgreen}{\mathsf{G}}
\newcommand{\sP}{\mathsf{P}}
\newcommand{\sH}{\mathsf{H}}

\newcommand{\RZ}{\mathbb{Z}}
\newcommand{\BVL}{{\Delta}_{\textrm{\tiny BV}}}
\newcommand{\CS}{\mathcal{S}}
\newcommand{\tte}{\mathtt{e}}
\newcommand{\ttc}{\mathtt{c}}

\newcommand{\ttF}{\mathtt{F}}
\newcommand{\e}{{\mathrm{e}}}
\newcommand{\FR}{\mathbbm{R}} 
\newcommand{\dd}{\mathrm{d}}

\numberwithin{equation}{section}

\title{BV quantization of $\phi^3$-theory on $\lambda$-Minkowski space: Tree-level correlation functions}
\ShortTitle{Quantum $\phi^3$-theory on $\lambda$-Minkowski space}

\author[a]{Dj. Bogdanovi\'c}
\author*[a]{M. Dimitrijevi\' c \'Ciri\'c}
\author[a]{S. Djordjevi\' c}
\author[b]{R. J. Szabo}

\affiliation[a]{University of Belgrade, Faculty of Physics\\
Studentski trg 12, Belgrade, Serbia}

\affiliation[b]{Department of Mathematics, Heriot--Watt University, Edinburgh, United
Kingdom\\
Maxwell Institute for
Mathematical Sciences, Edinburgh, United Kingdom}

\emailAdd{djbogdan@ipb.ac.rs}
\emailAdd{dmarija@ipb.ac.rs}
\emailAdd{stefan.djordjevic@ff.bg.ac.rs}
\emailAdd{R.J.Szabo@hw.ac.uk}

\abstract{We review the quantization of scalar field theory on  $\lambda$-Minkowski space using the  Batalin--Vilkovisky (BV) formalism. We consider $\phi^3$-theory in two different quantization schemes: standard and braided. While  standard BV quantization is based on an ordinary $L_\infty$-algebra,  braided BV quantization is based on a braided $L_\infty$-algebra. We compare the tree-level three-point and four-point correlation functions in the two approaches. For the four-point function,  standard quantization leads to two inequivalent classes of diagrams with different noncommutative contributions, whereas  braided quantization yields only a single class of diagrams with noncommutativity entering solely through an overall phase factor depending on the external momenta. }

\FullConference{Proceedings of the Corfu Summer Institute 2025 "School and Workshops on Elementary Particle Physics and Gravity" (CORFU2025)\\
27 April -- 28 September, 2025\\
Corfu, Greece\\}

\begin{document}
\maketitle

\section{Introduction}

A systematic way to construct noncommutative deformations of spacetime together with their deformed symmetry structures is provided by the Drinfel'd twist formalism \cite{Drinfeld}. In this approach, the twist deforms the Hopf algebra of spacetime symmetries by modifying its coproducts, and hence the Leibniz rule for the action of symmetry generators on products of fields. At the same time, it induces a deformation of the algebra of functions through a star-product. The twist formalism allows one to formulate noncommutative field theories with well-controlled twisted symmetries. Besides the canonical Moyal deformation, it also naturally accommodates other noncommutative spacetimes with  twisted Poincar\'e invariance \cite{vanTongeren}.

One of the most prominent characteristic features of noncommutative quantum field theories is the appearance of ultraviolet/infrared (UV/IR) mixing, whereby ultraviolet divergences reemerge as infrared singularities in loop diagrams \cite{Minwalla:1999px, U1Reviewa}. This phenomenon, first observed for scalar field theories on Moyal space, remains one of the central obstacles in the construction of renormalizable noncommutative quantum field theories.

In this  contribution we focus on  $\lambda$-Minkowski space, which arises from an angular Drinfel'd twist and provides an example of a  Lie algebraic noncommutative deformation of Minkowski space \cite{AngTwistRNBH}. Scalar field theory on this spacetime was previously studied using  standard Feynman quantization and the conventional plane wave basis for the Fourier decomposition of fields, where noncommutativity manifests itself through a deformed momentum composition law and leads to nontrivial quantum effects, including UV/IR mixing \cite{AngTwistMi}.

More recently, an alternative algebraic approach to quantization based on homotopy algebra methods and the Batalin--Vilkovisky (BV) formalism was developed for noncommutative field theories \cite{SzaboAlex, BraidedQED, BraidedScalarMW, BraidedScalarQED, UsBessel}. A key outcome of this framework is that the same classical noncommutative scalar theory can be encoded by two different homotopy algebraic structures~\cite{Giotopoulos:2021ieg}: an ordinary $L_\infty$-algebra, leading to  standard BV quantization, and a braided $L_\infty$-algebra, leading to  braided BV quantization. In  standard BV quantization, one recovers the usual noncommutative scalar field theory with  nonplanar diagrams and  UV/IR mixing. In  braided BV quantization, noncommutativity is incorporated directly into the braided homotopy algebraic structure, leading to the absence of nonplanar diagrams and therefore the absence of the UV/IR mixing. These two quantization schemes were investigated in detail for  $\phi^3$-theory on $\lambda$-Minkowski space in \cite{UsBessel}.

Here we present a brief review of this construction, concentrating on the simplest correlation functions that already capture the differences between the two quantization schemes. We discuss the two $L_\infty$-algebraic descriptions of the noncommutative $\phi^3$-theory, comparing the resulting tree-level three-point and four-point correlation functions. These examples  complement  the one-loop analysis of the two-point function from~\cite{UsBessel}. They provide a transparent illustration of how the standard and braided BV formalisms result in two different quantum field theories, even though they originate from the same classical action.

The rest of this contribution is organized as follows. In Section~\ref{sec:2} we review the two $L_\infty$-algebra descriptions of $\phi^3$-theory on $\lambda$-Minkowski space and explain how they lead to the standard and braided BV quantization schemes. In Section~\ref{sec:3} we compute the tree-level three-point and four-point correlation functions in both approaches, and discuss their properties. In Section~\ref{sec:4} we conclude by summarizing the main results and comment on possible directions for future work.

\section{$\boldsymbol{\phi^3}$-theory on $\boldsymbol\lambda$-Minkowski space}
\label{sec:2}

Consider four-dimensional Minkowski space $\mathbbm{R}^{1,3}$ with Cartesian coordinates $(t,x,y,z)$, where $t$ is time. For a deformation parameter $\lambda\in\FR$, the $\lambda$-Minkowski space is defined by the angular Drinfel'd twist \cite{AngTwistRNBH} 
\begin{equation} \label{TwistLambda}
{\cal F} = \exp \left(-\frac{\ii\, \lambda}{2}\, (\partial_z \otimes \partial_{\varphi} - \partial_{\varphi} \otimes \partial_z)\right) \ .
\end{equation}
The commuting vector fields $\partial_z$ and $\partial_{\varphi}=x\,\partial_y - y\,\partial_x$ are the generators of translations in the $z$-direction and rotations around the $z$-axis, respectively. 

The pointwise product between functions  $\mu (f \otimes g) = f \cdot g$ is replaced by the noncomutative star-product, defined via
\begin{align} 
\begin{split}
f \star g & = \mu \circ \CF^{-1} (f \otimes g) \\[4pt]
& = f\cdot g + \sum_{n=1}^\infty\,\bigg(\frac{\ii\,\lambda}2\bigg)^n \ \sum_{k=0}^n\,\frac{(-1)^k}{k!\,(n-k)!} \, \partial_z^{n-k}\,\partial_\varphi^k f\cdot\partial_{\varphi}^{n-k}\,\partial_z^k g \ .\label{star}
\end{split}
\end{align}
The braided commutativity of the star-product (\ref{star}) is expressed via the triangular $R$-matrix
\begin{align}
f \star g &= \mu\circ\CF^{-1}\,\RR^{-1} (g \otimes f) = \sfR_{\alpha}(g) \star \sfR^{\alpha}(f) \ ,
\end{align}
where
\begin{align}
\RR^{-1} &= \CF^2 =: \sfR_{\alpha} \otimes \sfR^{\alpha} \ .
\end{align}

The twist (\ref{TwistLambda}) deforms the standard  Poincar\'e Hopf algebra of symmetries into the twisted Poincar\'e Hopf algebra. In particular, the coproduct of translation generators, i.e. of  Cartesian momentum operators $P_\mu$, is deformed as
\begin{align}
\begin{split}
\sfDelta_{\cal F}(P_{t}) &= P_{t}\otimes 1 + 1\otimes P_{t} \ , \\[4pt]
\sfDelta_{\cal F}(P_{z}) &= P_{z}\otimes 1 + 1\otimes P_{z} \ , \\[4pt]
\sfDelta_{\cal F}(P_{x}) &= P_{x}\otimes \cos\big( \tfrac{\lambda}{2}\,P_z \big) + \cos\big(
\tfrac{\lambda}{2}\,P_z \big)\otimes P_{x} + P_{y}\otimes \sin\big( \tfrac{\lambda}{2}\,P_z \big) - \sin\big(
\tfrac{\lambda}{2}\,P_z
\big)\otimes P_{y} \ , \label{TwistedPoincareCoproduct}\\[4pt]
\sfDelta_{\cal F}(P_{y}) &= P_{y}\otimes \cos\big( \tfrac{\lambda}{2}\,P_z \big) + \cos\big(
\tfrac{\lambda}{2}\,P_z \big)\otimes P_{y} - P_{x}\otimes \sin\big( \tfrac{\lambda}{2}\,P_z \big) + \sin\big(
\tfrac{\lambda}{2}\,P_z
\big)\otimes P_{x} \ . 
\end{split}
\end{align}
We will see later that this deformation results in a deformation of the conservation of momentum in vertices and correlation functions.

Using the star-product (\ref{star}) instead of the pointwise product leads to the action for the standard noncommutative cubic scalar field theory on four-dimensional Minkowski space, which can be written as
\begin{equation}
S_{\rm cl} = \int_{\FR^4}\, \d^4 x \ \Big( \frac{1}{2}\,\phi\,\big(-\square - m^2\big)\,\phi - \frac{g}{3!}\,\phi\star\phi\star\phi \Big)\ . \label{S_class}
\end{equation}
In what follows  we show that this action functional is reproduced by two different cyclic $L_\infty$-algebras.

\subsection{Standard $\boldsymbol{L_\infty}$-algebra}

The underlying vector space of the standard $L_\infty$-algebra of  noncommutative scalar field theory with cubic interaction is the graded vector space $V = V_1 \oplus V_2$, consisting of the space of fields $\phi\in V_1=C^\infty(\FR^4)$ and the space of antifields $\phi^+\in V_2=C^\infty(\FR^4)$. This vector space is equipped with the pair of nontrivial brackets 
\begin{align}
\mu_1:V_1 \longrightarrow V_{2} \quad , \quad \mu_2:V_1 \otimes V_1 \longrightarrow V_{2} \ ,
\end{align}
acting on elements $\phi, \phi_1,\phi_2 \in V_1$ as
\begin{align}
\begin{split}
\mu_1(\phi) &= - (\Box + m^2)\, \phi \ , \\[4pt]
\mu_2(\phi_1,\phi_2) &= \frac g2\, (\phi_1 \star \phi_2 + \phi_2\star\phi_1) \ .\label{LBracketsStandard}
\end{split}
\end{align}
The binary bracket $\mu_2$ is strictly symmetric, $\mu_2(\phi_1,\phi_2) = \mu_2(\phi_2,\phi_1)$. The noncommutative deformation enters through the star-product between the fields $\phi_1$ and $\phi_2$. 

To formulate the action functional and the action principle, a strictly graded symmetric cyclic pairing of degree $-3$ is defined by the integral
\begin{equation}
\langle \phi, \phi ^+\rangle  = \int_{\FR^4}\, \d^4 x \ \phi \star \phi^+ = \int_{\FR^4}\, \d^4 x \ \phi \cdot \phi^+ \ . \label{Pairing}
\end{equation}

Using this data, we can write the classical action as the homotopy Maurer--Cartan functional
\begin{align}
\begin{split}
S_{\rm cl} & = \frac{1}{2!} \, \langle \phi, \mu_1(\phi)\rangle
- \frac{1}{3!} \, \langle\phi, \mu_2(\phi, \phi)\rangle  \\[4pt]
& = \int_{\FR^4}\, \d^4 x \ \Big( \frac{1}{2}\,\phi\,\big(-\square - m^2\big)\,\phi - \frac{g}{3!}\,\phi\star\phi\star\phi \Big) \ , \label{S_class_Standard}
\end{split}
\end{align}
and the corresponding equation of motion as the Maurer--Cartan equation
\begin{align}
    \begin{split}
F_\phi & = \mu_1(\phi) - \frac{1}{2!}\,\mu_2(\phi,\phi)  \\[4pt]
& = \big(-\square - m^2\big)\,\phi - \frac{g}{2}\,\phi\star\phi = 0  \label{EoM_Standard}
\end{split}
\end{align}
of the $L_\infty$-algebra.
As expected, the standard noncommutative action (\ref{S_class}) is recovered.

Next we choose the plane wave basis for the Fourier decomposition of fields and antifields in $V$ as
\begin{align}\label{BazisPl}
\begin{split}
\tte_{k}(x) &= \tte_{(E, k_x, k_y, k_z)}(t, x, y, z) := \e^{- \ii\, k\cdot x} \ \in \ V_1 \ , \\[4pt]
\tte^{k}(x) &= \tte_{(-E, -k_x, -k_y, -k_z)}(t, x, y, z) := \e^{\,\ii\, k\cdot x}\ \in \ V_2 \ .
\end{split}
\end{align}
These basis vectors are  dual with respect to the cyclic structure \eqref{Pairing}:
\begin{align}\nn
\langle \tte_{k_1}\,,\,\tte^{k_2}\rangle = \int_{\FR^4}\, \dd^4x \ \e^{-\ii\,(k_1-k_2)\cdot x} = (2\pi)^4 \ \delta(k_1-k_2) \ .
\end{align}

Following \cite{UsBessel} the corresponding interacting BV master action  is given by
\begin{align}\label{SintPl}
\CS^{\tt pl}_{\rm int} = \int_{k_1} \ \int_{k_2} \ \int_{k_3} \  V_{\tt pl}(k_1, k_2, k_3) \ \tte^{k_1}\odot \tte^{k_2} \odot \tte^{k_3}  \ ,
\end{align}
with the shorthand notation 
\begin{align} 
\int_{k} := \int_{\FR^4}\, \frac{\d^4 k}{(2\pi)^4} \ \ ,
\end{align}
and the vertex
\begin{align}\label{VerteksPl}
V_{\tt pl}(k_1, k_2, k_3) = -\frac{g}{3!} \ (2\pi)^4 \ \delta(k_1 +_\star k_2 +\star k_3) \ .
\end{align}

The deformed momentum conservation laws implied by \eqref{TwistedPoincareCoproduct} are implemented by the Dirac distribution
\begin{align}
\delta (k_1 +_\star \cdots +_\star k_n) =  \frac1{(2\pi)^4}\,\langle\tte_{k_1},\tte_{k_2}\star\cdots\star\tte_{k_n}\rangle = \int_{\FR^4}\, \frac{\d^4x}{(2\pi)^4} \  \tte_{k_1} \star \cdots \star \tte_{k_n} \ .\label{StarDelta3_2} 
\end{align}
By the cyclicity property of \eqref{Pairing}, it satisfies
\begin{align} \label{eq:delta-cyclic}
\delta (k_1 +_\star k_2 +_\star\cdots +_\star k_n)=\delta (k_2 +_\star  \cdots +_\star k_n +_\star k_1)
= \delta \big(k_1 +  (k_2 +_\star\cdots+_\star  k_n)\big) \ .
\end{align}
In particular, $\delta(k_1+_\star k_2) = \delta(k_1+k_2)$. Then it is clear that the vertex (\ref{VerteksPl}) is cyclic:
\begin{equation}
V_{\tt pl}(k_1, k_2, k_3) = V_{\tt pl}(k_2, k_3, k_1) = V_{\tt pl}(k_3, k_1, k_2) \ .\label{VPlCyclic}    
\end{equation}

Explicitly, the star-sums of two and three momenta read as~\cite{AngTwistMi}
\begin{align}
k+_\star p &= \sfLambda(p_z)\,k + \sfLambda(-k_z)\,p  \ ,\label{StarSumTwo_2}\\[4pt]
k +_{\star} p +_{\star} q &= \sfLambda(p_z + q_z)\,k + \sfLambda(-k_z + q_z)\,p  + \sfLambda(-k_z -p_z)\,q \ , \label{StarSumThree_2}
\end{align}
where the rotation matrix
\begin{equation}
\sfLambda (p_z) = {\small \left( \begin{matrix}
     1 & 0 & 0 & 0\\
     0 & \cos \big(\frac{\lambda}{2}\,p_z\big) & \sin \big(\frac{\lambda}{2}\,p_z\big) &0\\
     0 & -\sin \big(\frac{\lambda}{2}\,p_z\big) & \cos \big(\frac{\lambda}{2}\,p_z\big) &0\\
     0 & 0 & 0 & 1
    \end{matrix} \right) } \normalsize 
\end{equation}
implements a rotation through angle $\frac\lambda2\,p_z$ in the $(x,y)$-plane of momentum space. While the star-sum does not affect the components $(E,k_z)$, it modifies the addition law of the components $(k_x,k_y)$.

The plane wave basis (\ref{BazisPl}) is not a natural basis for the twist (\ref{TwistLambda}). Namely, the star-product of two plane waves is no longer a plane wave:
\begin{equation}
\e^{-\ii\, k_1\cdot x}\star \e^{-\ii\, k_2\cdot x} = \tte_{k_1 +_\star k_2}(x) = \e^{-\ii\,(k_1 +_\star k_2)\cdot x
} \ , \label{PlWavesProduct} 
\end{equation}
with $k_1 +_\star k_2$ defined in (\ref{StarSumTwo_2}). Since the BV quantization of $\phi^3_\star$-theory will be based on the standard $L_\infty$-algebra (\ref{LBracketsStandard}), there is no preferred basis and we can continue working with plane waves. This will change in the braided BV quantization of $\phi^3_\star$-theory.

To prepare all the necessary tools for calculating correlation functions, in addition to the interacting BV master action (\ref{SintPl}) we now calculate the tree-level two-point function, i.e. the propagator, in the BV formalism following~\cite{UsBessel}. Starting from the Poincar\'e-invariant Feynman propagator and its momentum space representation,
\begin{equation}
\sgreen = -\frac1{\square + m^2+\ii\,\epsilon} \quad , \quad \widetilde \sgreen(k) = \frac1{k^2-m^2-\ii \,\epsilon}  \ ,\label{TildeG}
\end{equation}
with $\epsilon\to0^+$ and $k^2 = E^2-k_x^2-k_y^2-k_z^2$, we define the contracting homotopy $\sfh:V_2\longrightarrow V_1$ by
\begin{align}
    \sfh (\tte^k) = -\widetilde \sgreen(k) \ \tte^k \ .
\end{align}
The tree-level two-point function  $\widetilde G_2(p_1, p_2)^{\swzero}$ follows as
\begin{align}
\begin{split}
\widetilde G_2(p_1, p_2)^{\swzero} &= -\ii\,\hbar\,\BVL\,\sH\,(\tte^{p_1} \odot \tte^{p_2}) \\[4pt]
&= -\ii\,\hbar\,\langle \tte^{p_1}\,,\sfh(\tte^{p_2})\rangle \\[4pt]
&= \ii\,\hbar\, (2\pi)^4 \ \delta(p_1 +_\star p_2) \  {\widetilde{\sgreen}}(p_1) \\[4pt]
&= \ii\,\hbar\, (2\pi)^4 \ \delta(p_1 + p_2) \  {\widetilde{\sgreen}}(p_1) \ ,  \label{C2Pl}
\end{split}
\end{align}
where $\BVL$ is the BV Laplacian and $\sH$ denotes the ``thickened'' extension of $\sfh$ to symmetric products of antifields, see~\cite{UsBessel} for details.

Using (\ref{SintPl}) and (\ref{C2Pl}), in Section~\ref{sec:3} we will calculate the tree-level three-point and four-point correlation functions.

\subsection{Braided $\boldsymbol{L_\infty}$-algebra}

As in the case of the standard $L_\infty$-algebra of $\phi^3_\star$-theory, the underlying braided $L_\infty$-algebra of the scalar field theory with cubic interaction is the graded vector space $V = V_1 \oplus V_2$, consisting of the space of fields $V_1=C^\infty(\FR^4)$ and the space of antifields $V_2=C^\infty(\FR^4)$. Again, the only nontrivial action of brackets is on fields
\begin{align}
\begin{split}
\mu_1^{\RR}(\phi) &= \mu_1(\phi) = - (\Box + m^2)\, \phi \ , \\[4pt]
\mu_2^{\RR}(\phi_1,\phi_2) &= g\, \phi_1 \star \phi_2 \ ,  \label{BraidedLBrackets}
\end{split}
\end{align}
but the binary bracket is now braided symmetric, $\mu_2^{\RR}(\phi_1,\phi_2) = \mu_2^\RR \big(\sfR_{\alpha} (\phi_2) , \sfR^{\alpha}(\phi_1) \big)$. More details on the construction and  properties of braided $L_\infty$-algebras can be found in \cite{BraidedLinf,Giotopoulos:2021ieg}.

The strictly graded symmetric pairing is the same as in (\ref{Pairing}), and the classical action and the corresponding equation of motion are the same as in (\ref{S_class_Standard}) and (\ref{EoM_Standard}).

However, the braided $L_\infty$-algebra formalism is adapted to the braided (deformed) underlying Poincar\'e symmetry, and the plane wave basis is no longer suitable since it does not diagonalize the twist (\ref{TwistLambda}): the star-product of two plane waves is no longer a plane wave, cf. (\ref{PlWavesProduct}). The basis adapted to the deformed Poincar\'e symmetry is written in cylindrical coordinates and is given by the cylindrical harmonics
\begin{align}
\begin{split}
\ttc_{k}(x) &= \ttc_{(E,\alpha,\ell,k_z)}(t,r,\varphi,z) := J_\ell(\alpha\, r) \ \e^{\,\ii\, \ell\, \varphi} \ \e^{-\ii\, E\, t+\ii\, k_z\, z} \ \in \ V_1 \ , \\[4pt]
\ttc^{k}(x) &= \ttc_{(-E,\alpha,-\ell,-k_z)}(t,r,\varphi,z) := J_\ell(\alpha\, r) \ \e^{-\ii\, \ell\, \varphi} \ \e^{\,\ii\, E\, t-\ii\, k_z\, z} \ \in \ V_2 \ , \label{BasisCyl}
\end{split}
\end{align}
where $J_\ell$ are the cylindrical Bessel functions.
The variables $k = (E,\alpha,\ell,k_z)$ have the following physical interpretation: $E\in\FR$ is the energy and $k_z\in\FR$ is the linear momentum in the $z$-direction, just as before. The variable $\alpha\in[0,\infty)$ is the magnitude of the radial momentum in the $(x,y)$-plane, and $\ell\in\RZ$ is the angular momentum about the $z$-axis, that is, an eigenvalue of the operator $P_\varphi=-\ii\,\partial_\varphi$.

Using the orthogonality relation for  cylindrical Bessel functions 
\begin{align}
\int_0^\infty\,r\,\dd r \ J_{\ell} (\alpha_1\, r) \, J_{-\ell} (\alpha_2\, r)  = \frac{(-1)^{\ell}}{\alpha_1} \ \delta (\alpha_1 - \alpha_2) \ , \label{2Bessel} 
\end{align}
it is easy to check that the basis vectors (\ref{BasisCyl}) are indeed dual:
\begin{align}\label{eq:cylpairing}
\begin{split}
\langle \ttc_{k_1} \, , \ttc^{k_2} \rangle &= \int_{\FR\times\FR}\, \d t \ \d z \ \int_0^\infty\, r \, \d r \ \int_0^{2\pi}\,\d \varphi \ J_{\ell_{1}}(\alpha_{1}\, r) \ \e^{\,\ii\, \ell_{1}\, \varphi} \ \e^{-\ii\, E_1\, t} \ \e^{\,\ii\, k_{1z}\, z} \\[-8pt]
& \hspace{6cm} \times \, J_{\ell_{2}}(\alpha_{2}\, r) \ \e^{- \ii\, \ell_{2}\, \varphi} \ \e^{\,\ii\, E_2\, t} \ \e^{- \ii\, k_{2z}\, z} \\[4pt]
& = (2\pi)^3\ \delta(E_1 - E_2)\ \delta(k_{1z} - k_{2z}) \ \delta_{\ell_{1} , \ell_{2}}\ \frac{\delta(\alpha_{1} - \alpha_{2})}{\alpha_{1}} \ .
\end{split}
\end{align}

Following \cite{UsBessel} the corresponding braided interacting BV master action is given by
\begin{align}
\CS_{\rm int}^{\RR} &= \int_{k_1}\hspace{-5mm}\mbox{$\sum$} \ \ \int_{k_2}\hspace{-5mm}\mbox{$\sum$} \ \ \int_{k_3}\hspace{-5mm}
\mbox{$\sum$} \ \  V_\RR(k_1, k_2, k_3) \ \ttc^{k_1}\odot_\RR \ttc^{k_2} \odot_\RR \ttc^{k_3}  \ ,  \label{SintCyl} 
\end{align}
where we use the shorthand notation 
\begin{align}
\int_{k}\hspace{-4mm}\mbox{$\sum$} \ :=\int_{\FR\times\FR} \, \frac{\d E\,\d z}{(2\pi)^2} \ \int_0^\infty\,\alpha\,\d \alpha \ \frac1{2\pi}\,\sum_{\ell\in\RZ} \ \ ,
\end{align}
and we introduced the braided symmetric tensor product $\odot_\RR = \odot\circ\CF^{-1}$ with
\begin{equation}
\ttc^{k_1}\odot_\RR \ttc^{k_2} = \sfR_{\alpha} (\ttc^{k_2})\odot_\RR \sfR^{\alpha} (\ttc^{k_1}) = \e^{\,\ii\,\lambda\,(k_{1z}\,\ell_2-k_{2z}\,\ell_1)} \ \ttc^{k_2}\odot_\RR\ttc^{k_1}  \ .
\end{equation}
The vertex $V_\RR(k_1, k_2, k_3)$ is defined as
\begin{align}
\begin{split} 
V_\RR(k_1, k_2, k_3) &= -\frac{g}{3!} \  \e^{\,\frac{\ii\, \lambda}{2}\, \sum\limits_{a<b}\, (k_{az}\, \ell_{b} - k_{bz}\, \ell_{a})} \  (2\pi)^3 \ \delta(E_1+E_2+E_3) \ \delta(k_{1z}+k_{2z}+k_{3z}) \\[-5pt]
& \hspace{6cm} \times \delta_{\ell_{1} + \ell_{2} + \ell_{3}, 0} \ \ttF_{\ell_{1}, \ell_{2}, \ell_{3}} (\alpha_{1}, \alpha_{2}, \alpha_{3})  \ . \label{VertexCyl}
\end{split}
\end{align}

The function $\ttF_{\ell_{1}, \ell_{2}, \ell_{3}} (\alpha_{1}, \alpha_{2}, \alpha_{3})$ is defined as an integral of the product of three cylindrical Bessel functions:
\begin{equation} \label{Besself}
\ttF_{\ell_{1}, \ell_{2}, \ell_{3}} (\alpha_{1}, \alpha_{2}, \alpha_{3}) :=  \int_0^\infty\, r \, \d r \ J_{\ell_{1}}(\alpha_{1}\, r) \ J_{\ell_{2}}(\alpha_{2}\, r) \ J_{\ell_{3}}(\alpha_{3}\, r) \ .
\end{equation}
This integral can be computed explicitly \cite{Gervois:1984, Jackson:1972} provided that $\ell_{1} + \ell_{2} + \ell_{3} = 0$, which is satisfied here due to angular momentum conservation, and it is nontrivial when the radial momenta satisfy the triangle condition. The interaction vertex is formally similar to the braided interaction vertex in the Moyal case~\cite{BraidedScalarMW}, due to the choice of  basis adapted to the symmetry of the twist. The vertex (\ref{VertexCyl}) is braided symmetric:
\begin{align}
V_\RR(k_1, k_2, k_3) &= \e^{\,\ii\, \lambda\, (k_{1z}\, \ell_{2} - k_{2z}\, \ell_{1})} \ V_\RR(k_2, k_1, k_3) = \e^{\,\ii\, \lambda\, (k_{2z}\, \ell_{3} - k_{3z}\, \ell_{2})} \ V_\RR(k_1,k_3,k_2) \ ,
\end{align}
 and cyclic:
 \begin{align}
V_\RR(k_1, k_2, k_3) &= V_\RR(k_2, k_3, k_1)\ .
\label{VertexCylSymmetries}
\end{align}

The cylindrical harmonics also diagonalize the Klein--Gordan operator and hence the Feynman propagator \eqref{TildeG},
\begin{align}
    \sfh(\ttc^k) = -\widetilde\sgreen(k) \ \ttc^k \ ,
\end{align}
with $k^2 = E^2-\alpha^2-p_{z}^2$.
The tree-level two-point function is now given by
\begin{align}
\begin{split} 
\widetilde C_2(p_1, p_2)^{\swzero} &= -\ii\,\hbar\,\BVL\,\sH\,(\ttc^{p_1} \odot \ttc^{p_2}) \\[4pt]
&= -\ii\,\hbar\,\langle \ttc^{p_1}\,,\sfh(\ttc^{p_2})\rangle \\[4pt]
&= \ii\,\hbar\, (2\pi)^3\, (-1)^{\ell_{1}} \ \delta(E_1 + E_2) \ \delta(p_{1z} + p_{2z}) \ \delta_{\ell_{1} + \ell_{2}, 0} \ \frac{\delta(\alpha_{1} - \alpha_{2})}{\alpha_{1}} \ {\widetilde{\sgreen}}(p_1) \ .  \label{C2Cyl}
\end{split}
\end{align}

\section{Tree-level correlation functions}
\label{sec:3}

In this contribution we focus on  tree-level correlation functions. We  present and compare results for the tree-level three-point and four-point  functions in two different BV  formalisms: the standard BV quantization and  braided BV quantization. These computations complement the one-loop calculations performed in \cite{UsBessel}, where more details can be found.

\subsection{Standard Batalin--Vilkovisky quantization}

Since the standard BV quantization is based on the standard $L_\infty$-algebra, there are no restrictions on the basis of fields we should use and we continue working with the plane wave basis, defined in~(\ref{BazisPl}).

\paragraph{Three-point function.}

We start from the definition of the tree-level three-point correlation function 
\begin{equation}\label{G3def}
\widetilde G_3(p_1,p_2, p_3)^\swzero =  \sP\,\big(-\ii\,\hbar\,\BVL\,\sH - \{\CS^{\tt pl} _{\rm int},-\}\,\sH\big)^3\, (\tte^{p_1}\odot \tte^{p_2}\odot \tte^{p_3}) \ , 
\end{equation}
where $\sP$ is the projection to $\mathbbm{C}$ and $\{-,-\}$ denotes the BV antibracket, see~\cite{UsBessel} for details. Extracting only one-particle irreducible contributions
\begin{equation}
\widetilde G_3(p_1, p_2,p_3)^{\swzero}_{\1PI} =  -(-\ii\,\hbar\,\BVL\,\sH)^2 \,\big\{ \CS^{\tt pl}_{\rm int},\sH\, (\tte^{p_1}\odot \tte^{p_2}\odot \tte^{p_3})\big\} \, \Big|_{\1PI} \ , \label{G3Tree}
\end{equation}
we obtain 
\begin{align}
\begin{split}
\widetilde G_3(p_1, p_2,p_3)^{\swzero}_{\1PI} &= (\ii\,\hbar)^2\, \Big(-\frac{g}{3!}\Big) \, (2\pi)^4 \ \delta(p_{1}+_\star p_{2} +_\star p_{3}) \\
& \hspace{0.7cm} \times  \widetilde{\sgreen}(p_1) \, \widetilde{\sgreen}(p_2) \, \widetilde{\sgreen}(p_3) \ . \label{G3TreeTypicalResult} 
\end{split}
\end{align}
The star-sum of momenta in the Dirac distribution is defined in (\ref{StarDelta3_2}) and it encodes the deformed conservation of the Cartesian momenta. In this basis, the noncommutative deformation enters differently compared to results with the Moyal deformation where it appears as a phase factor.

To better understand this deformed conservation law, we express plane waves as Fourier series in cylindrical harmonics,
\begin{align}\label{eq:planetocylAdd}
    \tte_{k} = \sum_{\ell_k\in\RZ} \, \ii^{\,\ell_k} \ \e^{-\ii\,\ell_k \, \vartheta_k} \ \ttc_{k} \ ,
\end{align}
with
\begin{align}
\begin{split}
    \tte_k(x) &= \tte_{(E_k\,,\,k_x\,,\,k_y\,,\,k_z)}(t,x,y,z) = \e^{-\ii \,E_k\,t +\ii\,k_x\,x+\ii\,k_y\,y+\ii\,k_z\,z} \ ,\\[4pt]
    \ttc_k(x) &= \ttc_{(E_k\,,\,\alpha_{k}\,,\,\ell_k\,,\,k_z)}(t,r,\varphi,z) = J_{\ell_k}(\alpha_k\, r) \ \e^{\,\ii\, \ell_k\, \varphi} \ \e^{-\ii\, E_k\, t+\ii\, k_z\, z} \ ,
    \end{split}
\end{align}
where $\alpha_k = \big(k_x^2+k_y^2\big)^{1/2}$ and $\vartheta_k = \tan^{-1}({k_y}/{k_x})$. Motivated by the relation (\ref{eq:planetocylAdd}) as well as explicit calculations at low multiplicity and loop level, in~\cite{UsBessel} it is conjectured that the all orders $n$-point correlation functions in these two bases are related by
\begin{align}
\widetilde G_n(p_1, \dots, p_n) &= \sum_{\ell_1, \dots, \ell_n\in\RZ} \ \ii^{\,\ell_1 +\dots +\ell_n} \ 
\e^{-\ii\,\ell_1\, \vartheta_{1}-\cdots-\ii\,\ell_n\, \vartheta_{n}} \ \widetilde C_n(p_1, \dots, p_n) \ . 
\end{align}
The correlation functions $\widetilde C_n(p_1, \dots, p_n)$ are calculated in a completely analogous way as the corresponding correlators $\widetilde G_n(p_1, \dots, p_n)$, but with insertions of the cylindrical harmonic  states (\ref{BasisCyl})  instead of the plane wave states (\ref{BazisPl})~\cite{UsBessel}.

Let us consider the correlator (\ref{G3TreeTypicalResult}) in the special case where $p_1 = (E_{1}, 0,0, p_{1z})$, i.e.\ for purely axial momentum $\alpha_1=0$. Starting from (\ref{G3TreeTypicalResult}) and 
\begin{align}
\begin{split}
\widetilde C_3(p_1, p_2, p_3)_\1PI^{\swzero} &= (\ii\,\hbar)^2 \, \frac{g}{3!} \ \widetilde{\sgreen}(p_1) \ \widetilde{\sgreen}(p_2) \ \widetilde{\sgreen}(p_3) \ \e^{\,\frac{\ii\, \lambda}{2} \, (p_{1z}\, \ell_{2} - p_{2z}\, \ell_{1})} \ \delta_{\ell_{1}+\ell_{2}+\ell_{3},0}   \\
&\hspace{1cm} \times (2\pi)^3 \ \delta(E_{1}+E_{2}+E_{3}) \ \delta(p_{1z}+p_{2z}+p_{3z}) \ 
\ttF_{\ell_{1},\ell_{2},\ell_{3}} (\alpha_{1}, \alpha_{2}, \alpha_{3}) \ , \label{C3Example}
\end{split}
\end{align}
we obtain
\begin{align}
\begin{split}
\widetilde G_3(p_1, p_2, p_3)_\1PI^{\swzero}\big|_{\alpha_1=0} & =
\sum_{\ell_1, \ell_2, \ell_3\in\RZ} \ \ii^{\,\ell_1 +\ell_2 +\ell_3} \ 
\e^{-\ii\,\ell_1\, \vartheta_{1}-\ii\ell_2\, \vartheta_{2}-\ii\,\ell_3\, \vartheta_{3}} \ \widetilde C_3(p_1, p_2, p_3)_\1PI^\swzero\big|_{\alpha_1=0} \\[4pt]
& = (\ii\,\hbar)^2\,\frac{g}{3!} \ \widetilde{\sgreen}(p_1)  \ \widetilde{\sgreen}(p_2) \ \widetilde{\sgreen}(p_3) \, (2\pi)^3 \ \delta(E_{1}+E_{2}+E_{3}) \\
&\qquad \times  \delta(p_{1z}+p_{2z}+p_{3z}) \ \frac{\delta(\alpha_{2}-\alpha_{3})}{\alpha_{2}} \ \delta\big(\vartheta_{2} - \vartheta_{3} - \pi + \tfrac{\lambda}{2}\,p_{1z} \big) \ .
\end{split}
\end{align}
We conclude that the deformed conservation of momentum encoded by the Dirac distribution \smash{$\delta(p_1+_\star p_2+_\star p_3)\big|_{\alpha_1=0}$} can be interpreted as a modification of momentum conservation in the $(p_x,p_y)$-plane, namely as a shift of the relative polar angle between the momenta by $\frac{\lambda}{2}\,p_{1z}$. In this way, the noncommutative deformation modifies the momentum conservation law.

\paragraph{Four-point function.}

The tree-level four-point correlation function is defined as
\begin{equation}\label{G4def}
\widetilde G_4(p_1,p_2, p_3, p_4)^\swzero =  \sP\,\big(-\ii\,\hbar\,\BVL\,\sH - \{\CS^{\tt pl} _{\rm int},-\}\,\sH\big)^5\, (\tte^{p_1}\odot \tte^{p_2}\odot \tte^{p_3}\odot \tte^{p_4}) \ . 
\end{equation}
The one-particle irreducible contribution can be found in terms of the form
\begin{align}\label{G4TreeTypical}
\begin{split}
& \widetilde G_4(p_1,p_2, p_3, p_4)^\swzero_{\mbox{\tiny 1PI}} \\[4pt] & \qquad = (-\ii\,\hbar)^3\, (\BVL\,\sH)^2 \,  \{\CS^{\tt pl} _{\rm int},-\}\,\sH\, \BVL\,\sH \, \{\CS^{\tt pl} _{\rm int},-\}\,\sH \,(\tte^{p_1}\odot\tte^{p_2}\odot\tte^{p_3}\odot\tte^{p_4}) \, \big|_{\1PI} \\
& \hspace{2cm}+ (-\ii\,\hbar\, \BVL\,\sH)^3\, \{\CS^{\tt pl} _{\rm int},-\}\,\sH\, \{\CS^{\tt pl} _{\rm int},-\}\,\sH \,(\tte^{p_1}\odot\tte^{p_2}\odot\tte^{p_3}\odot\tte^{p_4}) \, \big|_{\1PI}  \ .
\end{split}
\end{align}
Performing a long but straightforward calculation, the final result contains three sets of diagrams corresponding to the all three $s$-, $t$-  and  $u $-channels. 

The $s$-channel is
\begin{align} 
\begin{split}
\widetilde G_4(p_1,p_2, p_3, p_4)_{\mbox{\tiny 1PI}}^{\swzero s} &= (-\ii\,\hbar)^3 \, \left(\frac{g}{3!}\right)^2 \ {\widetilde{\sgreen}}(p_1) \ {\widetilde{\sgreen}}(p_2) \ {\widetilde{\sgreen}}(p_3) \ {\widetilde{\sgreen}}(p_4) \ {\widetilde{\sgreen}}\big((-p_2) +_\star (-p_1)\big) \label{DDSDSstandardS}\\
& \hspace{1cm} \times (2\pi)^4 \ \Big(\delta\big((-p_3) +_\star (-p_4) +_\star (-p_2) +_\star (-p_1)\big) \\
& \hspace{3cm} +\, \delta\big((-p_4) +_\star (-p_3) +_\star (-p_2) +_\star (-p_1)\big) \Big) \ .
\end{split}
\end{align}
It can be represented diagrammatically as the sum of two inequivalent contributions due to a nontrivial and noncommuting momentum conservation law, with all momenta treated as incoming:
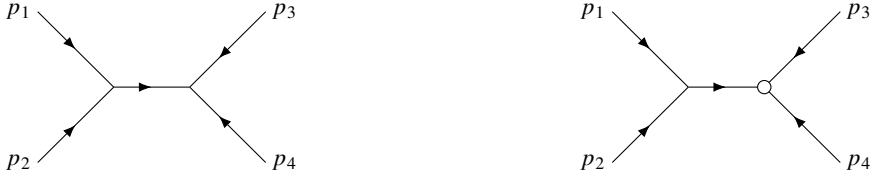
\begin{align*}
\begin{split}
\begin{minipage}{0.5\textwidth}
\centering
\footnotesize
\begin{tikzpicture}[scale=0.5, baseline]
    \coordinate (L) at (0,0);
    \coordinate (R) at (2,0);
    \draw[decoration={markings, mark=at position 0.5 with {\arrow{Latex}}}, postaction={decorate}] (-2,2) -- (L);
    \draw[decoration={markings, mark=at position 0.5 with {\arrow{Latex}}}, postaction={decorate}] (-2,-2) -- (L);
    \draw[decoration={markings, mark=at position 0.5 with {\arrow{Latex}}}, postaction={decorate}] (L) -- (R);
    \draw[decoration={markings, mark=at position 0.5 with {\arrow{Latex[reversed]}}}, postaction={decorate}] (R) -- (4,2);
    \draw[decoration={markings, mark=at position 0.5 with {\arrow{Latex[reversed]}}}, postaction={decorate}] (R) -- (4,-2);
    \node at (-2.5,2) {$p_1$};
    \node at (-2.5,-2) {$p_2$};
    \node at (4.5,2) {$p_3$};
    \node at (4.5,-2) {$p_4$};
\end{tikzpicture}
\end{minipage}
\hfill
\begin{minipage}{0.5\textwidth}
\centering
\footnotesize
\begin{tikzpicture}[scale=0.5, baseline]
    \coordinate (L) at (0,0);
    \coordinate (R) at (2,0);
    \draw[decoration={markings, mark=at position 0.5 with {\arrow{Latex}}}, postaction={decorate}] (-2,2) -- (L);
    \draw[decoration={markings, mark=at position 0.5 with {\arrow{Latex}}}, postaction={decorate}] (-2,-2) -- (L);
    \draw[decoration={markings, mark=at position 0.5 with {\arrow{Latex}}}, postaction={decorate}] (L) -- (R);
    \draw[decoration={markings, mark=at position 0.5 with {\arrow{Latex[reversed]}}}, postaction={decorate}] (R) -- (4,2);
    \draw[decoration={markings, mark=at position 0.5 with {\arrow{Latex[reversed]}}}, postaction={decorate}] (R) -- (4,-2);
    \node at (-2.5,2) {$p_1$};
    \node at (-2.5,-2) {$p_2$};
    \node at (4.5,2) {$p_3$};
    \node at (4.5,-2) {$p_4$};
    \fill[white] (R) circle (5pt);
    \draw (R) circle (5pt);
\end{tikzpicture}
\normalsize
\end{minipage}
\end{split}
\end{align*}
The second diagram represents the same formal expression as the first diagram but with one crucial difference: the permutation $p_3 \leftrightarrow p_4$ indicated by the white circle.

The $t$-channel has the form
\begin{align}
\begin{split}
\widetilde G_4(p_1,p_2, p_3, p_4)_{\mbox{\tiny 1PI}}^{\swzero t} &= (-\ii\,\hbar)^3 \, \left(\frac{g}{3!}\right)^2 \ {\widetilde{\sgreen}}(p_1) \ {\widetilde{\sgreen}}(p_2) \ {\widetilde{\sgreen}}(p_3) \ {\widetilde{\sgreen}}(p_4) \ {\widetilde{\sgreen}}\big((-p_3) +_\star (-p_1)\big)\label{DDSDSstandardT}\\
& \hspace{1cm} \times (2\pi)^4 \ \Big(\delta\big((-p_2) +_\star (-p_4) +_\star (-p_3) +_\star (-p_1)\big) \\
& \hspace{3cm} +\, \delta\big((-p_4) +_\star (-p_2) +_\star (-p_3) +_\star (-p_1)\big) \Big) \ ,
\end{split}
\end{align}
and is represented diagrammatically as
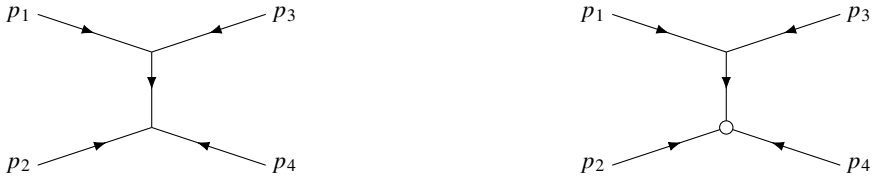
\begin{align*}
\begin{minipage}{0.5\textwidth}
\centering
\footnotesize
\begin{tikzpicture}[scale=0.5, baseline]
    \coordinate (T) at (1,1);
    \coordinate (B) at (1,-1);
    \draw[decoration={markings, mark=at position 0.5 with {\arrow{Latex}}}, postaction={decorate}] (-2,2) -- (T);
    \draw[decoration={markings, mark=at position 0.5 with {\arrow{Latex}}}, postaction={decorate}] (4,2) -- (T);
    \draw[decoration={markings, mark=at position 0.5 with {\arrow{Latex}}}, postaction={decorate}] (T) -- (B);
    \draw[decoration={markings, mark=at position 0.5 with {\arrow{Latex[reversed]}}}, postaction={decorate}] (B) -- (-2,-2);
    \draw[decoration={markings, mark=at position 0.5 with {\arrow{Latex[reversed]}}}, postaction={decorate}] (B) -- (4,-2);
    \node at (-2.5,2) {$p_1$};
    \node at (-2.5,-2) {$p_2$};
    \node at (4.5,2) {$p_3$};
    \node at (4.5,-2) {$p_4$};
\end{tikzpicture}
\end{minipage}
\hfill
\begin{minipage}{0.5\textwidth}
\centering
\footnotesize
\begin{tikzpicture}[scale=0.5, baseline]
    \coordinate (T) at (1,1);
    \coordinate (B) at (1,-1);
    \draw[decoration={markings, mark=at position 0.5 with {\arrow{Latex}}}, postaction={decorate}] (-2,2) -- (T);
    \draw[decoration={markings, mark=at position 0.5 with {\arrow{Latex}}}, postaction={decorate}] (4,2) -- (T);
    \draw[decoration={markings, mark=at position 0.5 with {\arrow{Latex}}}, postaction={decorate}] (T) -- (B);
    \draw[decoration={markings, mark=at position 0.5 with {\arrow{Latex[reversed]}}}, postaction={decorate}] (B) -- (-2,-2);
    \draw[decoration={markings, mark=at position 0.5 with {\arrow{Latex[reversed]}}}, postaction={decorate}] (B) -- (4,-2);
    \node at (-2.5,2) {$p_1$};
    \node at (-2.5,-2) {$p_2$};
    \node at (4.5,2) {$p_3$};
    \node at (4.5,-2) {$p_4$};
    \fill[white] (B) circle (5pt);
    \draw (B) circle (5pt);
\end{tikzpicture}
\normalsize
\end{minipage}
\end{align*}
with the two inequivalent contributions differing in the permutation $p_2 \leftrightarrow p_4$, indicated by the white circle.

The last channel, the $u$-channel, has the form
\begin{align}
\begin{split}
\widetilde G_4(p_1,p_2, p_3, p_4)_{\mbox{\tiny 1PI}}^{\swzero u} &= (-\ii\,\hbar)^3 \,\left(\frac{g}{3!}\right)^2 \ {\widetilde{\sgreen}}(p_1) \ {\widetilde{\sgreen}}(p_2) \ {\widetilde{\sgreen}}(p_3) \ {\widetilde{\sgreen}}(p_4) \ {\widetilde{\sgreen}}\big((-p_4) +_\star (-p_1)\big)\label{DDSDSstandardU}\\
& \hspace{1cm} \times (2\pi)^4 \ \Big(\delta\big((-p_2) +_\star (-p_3) +_\star (-p_4) +_\star (-p_1)\big) \\
& \hspace{3cm} +\, \delta\big((-p_3) +_\star (-p_2) +_\star (-p_4) +_\star (-p_1)\big) \Big) \ ,
\end{split}
\end{align}
and can be represented as
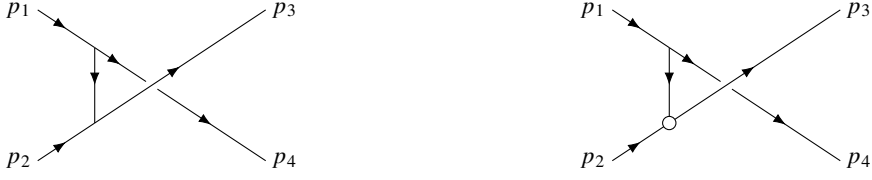
\begin{align*}
\begin{minipage}{0.5\textwidth}
\centering
\footnotesize
\begin{tikzpicture}[scale=0.5, baseline]
    \coordinate (T) at (-0.5,1);
    \coordinate (B) at (-0.5,-1);
    \coordinate (C) at (1,0);
    \draw[decoration={markings, mark=at position 0.5 with {\arrow{Latex}}}, postaction={decorate}] (-2,2) -- (T);
    \draw[decoration={markings, mark=at position 0.5 with {\arrow{Latex}}}, postaction={decorate}] (-2,-2) -- (B);
    \draw[decoration={markings, mark=at position 0.5 with {\arrow{Latex}}}, postaction={decorate}] (T) -- ($(C)+(-0.15,0.1)$);
    \draw[decoration={markings, mark=at position 0.5 with {\arrow{Latex}}}, postaction={decorate}] ($(C)+(+0.15,-0.1)$) -- (4,-2);
    \draw[decoration={markings, mark=at position 0.5 with {\arrow{Latex}}}, postaction={decorate}] (B) -- (4,2);
    \draw[decoration={markings, mark=at position 0.5 with {\arrow{Latex}}}, postaction={decorate}] (T) -- (B);
    \node at (-2.5,2) {$p_1$};
    \node at (-2.5,-2) {$p_2$};
    \node at (4.5,2) {$p_3$};
    \node at (4.5,-2) {$p_4$};
\end{tikzpicture}
\end{minipage}
\hfill
\begin{minipage}{0.5\textwidth}
\centering
\footnotesize
\begin{tikzpicture}[scale=0.5, baseline]
    \coordinate (T) at (-0.5,1);
    \coordinate (B) at (-0.5,-1);
    \coordinate (C) at (1,0);
    \draw[decoration={markings, mark=at position 0.5 with {\arrow{Latex}}}, postaction={decorate}] (-2,2) -- (T);
    \draw[decoration={markings, mark=at position 0.5 with {\arrow{Latex}}}, postaction={decorate}] (-2,-2) -- (B);
    \draw[decoration={markings, mark=at position 0.5 with {\arrow{Latex}}}, postaction={decorate}] (T) -- ($(C)+(-0.15,0.1)$);
    \draw[decoration={markings, mark=at position 0.5 with {\arrow{Latex}}}, postaction={decorate}] ($(C)+(+0.15,-0.1)$) -- (4,-2);
    \draw[decoration={markings, mark=at position 0.5 with {\arrow{Latex}}}, postaction={decorate}] (B) -- (4,2);
    \draw[decoration={markings, mark=at position 0.5 with {\arrow{Latex}}}, postaction={decorate}] (T) -- (B);
    \node at (-2.5,2) {$p_1$};
    \node at (-2.5,-2) {$p_2$};
    \node at (4.5,2) {$p_3$};
    \node at (4.5,-2) {$p_4$};
    \fill[white] (B) circle (5pt);
    \draw (B) circle (5pt);
\end{tikzpicture}
\normalsize
\end{minipage}
\end{align*}
with the two inequivalent contributions differing in the permutation $p_2 \leftrightarrow p_3$, indicated by the white circle.

Due to a deformation of momentum conservation enforced by the Dirac distributions of star-sums of momenta, each of the three channels (\ref{DDSDSstandardS}), (\ref{DDSDSstandardT}) and (\ref{DDSDSstandardU}) splits into two distinct contributions. While the internal propagator momenta at tree-level coincide for each contribution, the associated momentum conservation constraints differ by permutation of neighbouring external momenta. At the loop level, this feature manifests as the appearance of both planar and non-planar diagrams, as well as the emergence of UV/IR mixing \cite{UsBessel}.
 We shall now present an analysis of the quantitative difference between these {tree-level} contributions.

Since all Dirac distributions are formally related by relabeling of external momenta, it is sufficient to focus on a single pair of representative distributions, for example in (\ref{DDSDSstandardS}). Given that the propagating momentum is the star-sum of momenta $p_1$ and $p_2$, we specialize to the case in which these momenta are purely axial:
\begin{equation}
p_1 = (E_1,0,0,p_{1z}) \quad , \quad p_2 = (E_2,0,0,p_{2z}) \ .
\end{equation}

Following (\ref{StarSumThree_2}) and using cyclicity \eqref{eq:delta-cyclic}, for the first distribution in \eqref{DDSDSstandardS} we find
\begin{align}\label{eq:deltaU}
\begin{split}
    & \delta\big((-p_4) +_\star (-p_3) +_\star (-p_2) +_\star (-p_1)\big)\Big|_{\alpha_1=\alpha_2=0} \\[4pt]
    & \hspace{1cm} = \delta(E_1 + E_2 + E_3 + E_4) \  \delta(p_{1z} + p_{2z} + p_{3z} + p_{4z}) \\
    &\hspace{2cm} \times \delta\Big(p_{4x} + p_{3x}\cos\big(\tfrac{\lambda}{2}\,(p_{1z} + p_{2z})\big) - p_{3y}\sin\big(\tfrac{\lambda}{2}\,(p_{1z} + p_{2z})\big)\Big)\\
    &\hspace{3cm} \times \delta\Big(p_{4y} + p_{3x}\sin\big(\tfrac{\lambda}{2}\,(p_{1z} + p_{2z})\big) + p_{3y}\cos\big(\tfrac{\lambda}{2}\,(p_{1z} + p_{2z})\big)\Big) \ . 
    \end{split}
\end{align}
The second distribution in \eqref{DDSDSstandardS} is obtained from \eqref{eq:deltaU} by interchanging $p_3\leftrightarrow p_4.$

Momentum conservation in the transverse plane implies that the $(x,y)$-components of  the momenta $p_3$ and $p_4$ are related by a rotation. For the first distribution \eqref{eq:deltaU} this relation reads
\begin{align}
{\small\left(\begin{matrix}
     p_{4x}\\
     p_{4y}
    \end{matrix} \right) \normalsize} =  {\small \left( \begin{matrix}
    -\cos\big(\tfrac{\lambda}{2}\,(p_{1z} + p_{2z})\big) & \sin\big(\tfrac{\lambda}{2}\,(p_{1z} + p_{2z})\big)\\[2pt]
    -\sin\big(\tfrac{\lambda}{2}\,(p_{1z} + p_{2z})\big) & -\cos\big(\tfrac{\lambda}{2}\,(p_{1z} + p_{2z})\big)
    \end{matrix} \right) \, \left( \begin{matrix}
    p_{3x}\\
    p_{3y}
    \end{matrix} \right) \normalsize} \ .
\end{align}
The relation for the second distribution is analogous and is obtained from $p_3\leftrightarrow p_4$. Although the relative angle is identical in both cases, the orientation differs: in one contribution $p_3$ and $p_4$ together with the axial direction define a left-handed system, whereas in the other contribution they define a right-handed system.

As a result of the rotation, the magnitudes of their radial momenta coincide, $\alpha_3=\alpha_4$. Unlike the standard kinematic configuration at $\lambda=0$, where the transverse momenta are antiparallel, for $\lambda\neq0$ the angle between them deviates from $\pi$ by $\frac\lambda2\,(p_{1z}+p_{2z})$.

\subsection{Braided Batalin--Vilkovisky quantization}

Braided BV quantization was formulated in \cite{SzaboAlex} and was subsesquently applied to various scalar and spinor field theories with the Moyal deformation \cite{BraidedScalarMW, BraidedQED,BraidedScalarQED}. A detailed analysis of the braided quantization of cubic scalar field theory on $\lambda$-Minkowski space was performed in \cite{UsBessel}. In the following we focus on the tree-level three-point and four-point functions. Since braided BV quantization selects a preferred basis, compatible with the angular twist, in this subsection we work with the basis of cylindrical harmonics.

\paragraph{Three-point function.}

Once again we start from the definition of the one-particle irreducible contribution
\begin{equation}
\widetilde C_3^\RR(p_1, p_2,p_3)^{\swzero}_{\1PI} =  -(-\ii\,\hbar\,\BVL\,\sH)^2 \,\big\{ \CS _{\rm int}^\RR,\sH\, (\ttc^{p_1}\odot_\RR \ttc^{p_2}\odot_\RR\ttc^{p_3})\big\}_\RR \,\Big|_{\1PI} \ , \label{C3Tree}
\end{equation}
where $\{-\}_\RR = \{-\}\circ\CF^{-1}$ denotes the braided BV antibracket.
With the help of the Braided Wick Theorem~\cite{BraidedQED,BraidedScalarMW} they are of the form
\begin{align}
\begin{split}
\widetilde C_3^\RR(p_1, p_2,p_3)^{\swzero}_{\1PI} &= -(-\ii\,\hbar)^2 \, \int_{k_1}\hspace{-5mm}\mbox{$\sum$} \ \ \int_{k_2}\hspace{-5mm}\mbox{$\sum$} \ \ \int_{k_3}\hspace{-5mm}\mbox{$\sum$} \ \  V_\RR(k_1,k_2,k_3) \\
& \hspace{3cm} \times 
\langle \ttc^{k_3}, \sfh(\ttc^{p_1})\rangle \ \langle \ttc^{k_2}, \sfh(\ttc^{p_2})\rangle \ 
\langle \ttc^{k_1}, \sfh(\ttc^{p_3})\rangle  \\[4pt]
&= (-\ii\,\hbar)^2\, \Big(-\frac{g}{3!}\Big) \, (2\pi)^3 \  \delta(E_{1}+E_{2}+E_{3}) \ \delta(p_{1z}+p_{2z} +p_{3z} ) \ \delta_{\ell_{1}+\ell_{2}+\ell_{3}, 0}  \\
& \hspace{3cm} \times  \widetilde{\sgreen}(p_1) \, \widetilde{\sgreen}(p_2) \, \widetilde{\sgreen}(p_3) \ \e^{\frac{\ii\, \lambda}{2}\, \sum\limits_{a<b}\, (p_{az}\, \ell_b - p_{bz}\, \ell_a)}  \\
& \hspace{6cm} \times \ttF_{\ell_{1},\ell_{2},\ell_{3}} (\alpha_{1}, \alpha_{2}, \alpha_{3})  \ .\label{C3TreeTypical}
\end{split}
\end{align}

Provided that $\vec{\alpha}_{1}  + \vec{\alpha}_{2} + \vec{\alpha}_{3} = \vec 0$ and $\vec\alpha_i\neq\vec0$, where $\vec\alpha_i$ denotes the radial momentum vector with magnitude $\alpha_i$, the integral of three cylindrical Bessel functions reduces to
\begin{align}
\delta_{\ell_{1} + \ell_{2}+\ell_{3},0} \ \ttF_{\ell_{1},\ell_{2},\ell_{3}} (\alpha_{1}, \alpha_{2}, \alpha_{3}) = \delta_{\ell_{1} + \ell_{2}+\ell_{3}, 0} \,(-1)^{\ell_{3}}\,\frac{\cos \big(\ell_{1}\,\upsilon_{2} - \ell_{2}\,\upsilon_{1})}{\pi \,\alpha_{1}\,\alpha_{2} \sin (\upsilon_{3})} 
\end{align}
with the angles $\upsilon_{i}$ defined in the triangle
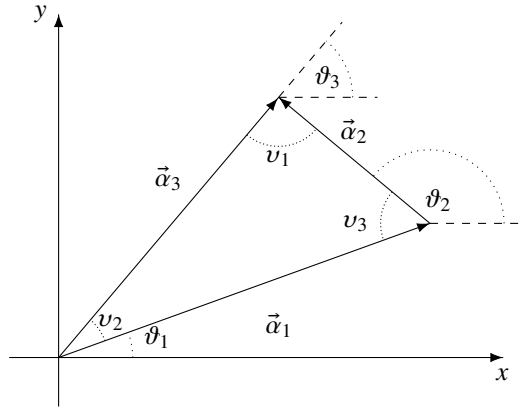
\begin{align}
\begin{split}
\small
    \begin{tikzpicture}[scale=0.65, baseline]
        \coordinate (O) at (0,0);
        \coordinate (A) at ($(O)+(20:8)$);
        \coordinate (B) at ($(A)+(140:4)$);
        \coordinate (S1) at ($0.5*(A)$);
        \coordinate (S2) at ($0.5*(A)+0.5*(B)$);
        \coordinate (S3) at ($0.5*(B)$);
        \draw[decoration={markings, mark=at position 1 with {\arrow[scale=1]{Latex}} }, postaction={decorate}] ($(O) +(180:1)$) -- +(0:10) node[below=0.5]{$x$};
        \draw[decoration={markings, mark=at position 1 with {\arrow[scale=1]{Latex}} }, postaction={decorate}] ($(O)+(-90:1)$) -- +(90:8) node[left=0.5]{$y$};
\draw[decoration={markings, mark=at position 1 with {\arrow[scale=1]{Latex}} }, postaction={decorate}] (O) -- (A);
        \draw[decoration={markings, mark=at position 1 with {\arrow[scale=1]{Latex}} }, postaction={decorate}] (A) -- (B);
        \draw[decoration={markings, mark=at position 1 with {\arrow[scale=1]{Latex}} }, postaction={decorate}] (O) -- (B);
\draw ($(S1)+(-45:1)$) node{$\Vec{\alpha}_{1}$};
        \draw ($(S2)+(90:0.6)$) node{$\Vec{\alpha}_{2}$};
        \draw ($(S3)+(90:1)$) node{$\Vec{\alpha}_{3}$};
        \draw ($(O)+(0:2)$) node[above]{$\vartheta_1$};
        \draw ($(O)+(45:1.5)$) node[below]{$\upsilon_2$};
        \draw ($(A)+(70:0.5)$) node{$\vartheta_2$};
        \draw ($(A)+(180:1.5)$) node{$\upsilon_3$};
        \draw ($(B)+(0:1)$) node[above]{$\vartheta_3$};
        \draw ($(B)+(-90:1.25)$) node{$\upsilon_1$};
\draw[dashed] (A) -- +(0:2);
        \draw[dotted] (A)+(0:1.5) arc [start angle=0, end angle=140, radius=1.5];
        \draw[densely dotted] (A)+(140:1) arc [start angle=140, end angle=200, radius=1];
        \draw[dashed] (B) -- +(0:2);
        \draw[dashed] (B) -- +(50:2);
        \draw[dotted] (B)+(0:1.5) arc [start angle=0, end angle=50, radius=1.5];
        \draw[densely dotted] (B)+(230:1) arc [start angle=230, end angle=320, radius=1];
        \draw[dotted] (O)+(0:1.5) arc [start angle=0, end angle=20, radius=1.5];
        \draw[densely dotted] (O)+(20:1) arc [start angle=20, end angle=50, radius=1];
    \end{tikzpicture}
    \nn
\normalsize
\end{split}
\end{align}

The final result for (\ref{C3TreeTypical}) is given by
\begin{align}
\begin{split}
& \widetilde C_3^\RR(p_1, p_2,p_3)^{\swzero}_{\1PI} \\[4pt]
& \hspace{1cm} = (-\ii\,\hbar)^2\, \Big(-\frac{g}{3!}\Big) \, (2\pi)^3 \ \delta(E_{1}+E_{2}+E_{3}) \ \delta(p_{1z}+p_{2z} +p_{3z} ) \ \delta_{\ell_{1} + \ell_{2} + \ell_{3}, 0}  \\
& \hspace{2cm} \times  \widetilde{\sgreen}(p_1) \ \widetilde{\sgreen}(p_2) \ \widetilde{\sgreen}(p_3) \ \e^{-\frac{\ii \, \lambda}{2}\, (p_{1z}\, \ell_{2} - p_{2z}\, \ell_{1} )} \, (-1)^{\ell_{3}}\,\frac{\cos \big(\ell_{1}\,\upsilon_{2} - \ell_{2}\,\upsilon_{1})}{\pi \,\alpha_{1}\,\alpha_{2} \sin (\upsilon_{3})} \ . \label{C3TreeTypicalResult2}
\end{split}
\end{align}
This result is qualitatively similar to the corresponding three-point function on Moyal space~\cite{BraidedScalarMW}: the noncommutative contribution appears only as a phase factor in the external momenta. Moreover, in braided BV quantization, this also holds at one-loop~\cite{UsBessel}. There is no mixing of external and internal momenta, and no non-planar diagrams, due to cancellations arising from the combination of the noncommutative interaction with the Braided Wick Theorem.

\paragraph{Four-point function.}

The four-point correlation function is defined as
\begin{equation}\label{C4def}
\widetilde C^{\RR}_n(p_1,p_2, p_3, p_4)^\swzero =  \sP\,\big(-\ii\,\hbar\,\BVL\,\sH - \{\CS^{\RR} _{\rm int},-\}_{\RR}\,\sH\big)^5\, (\ttc^{p_1}\odot_\RR\ttc^{p_2}\odot_\RR\ttc^{p_3}\odot_\RR\ttc^{p_4}) \ .
\end{equation}
Among all the terms arising in \eqref{C4def}, only two contain one-particle irreducible contributions
\begin{align}\label{C4TreeTypical}
\begin{split}
& \widetilde C^{\RR}_n(p_1,p_2, p_3, p_4)^\swzero_{\mbox{\tiny 1PI}} \\[4pt] & \quad = (-\ii\,\hbar)^3\, (\BVL\,\sH)^2 \, \big( \{\CS^{\RR} _{\rm int},-\}_{\RR}\,\sH \, (\BVL\,\sH) \, \{\CS^{\RR} _{\rm int},-\}_{\RR}\,\sH \,(\ttc^{p_1}\odot_\RR\ttc^{p_2}\odot_\RR\ttc^{p_3}\odot_\RR\ttc^{p_4}) \, \big|_{\1PI} \\
& \hspace{4cm} + \{\CS^{\RR} _{\rm int},-\}_{\RR}\,\sH \, \{\CS^{\RR} _{\rm int},-\}_{\RR}\,\sH \,(\ttc^{p_1}\odot_\RR\ttc^{p_2}\odot_\RR\ttc^{p_3}\odot_\RR\ttc^{p_4}) \, \big|_{\1PI}\,\big) \ .
\end{split}
\end{align}

Performing a long but straightforward calculation, the final result once again contains all three $s$-, $t$- and $u$-channels. As opposed to (\ref{DDSDSstandardS}), (\ref{DDSDSstandardT}) and (\ref{DDSDSstandardU}) in the standard theory, here a unique phase factor multiplies all three channel contributions and (up to an overall numerical factor) we find {the one-particle irreducible contribution
\begin{align} \label{DDSDSbraided}
\begin{split}
& \widetilde C_n(p_1,p_2, p_3, p_4)_{\mbox{\tiny 1PI}}^{\swzero} \\[4pt] & \hspace{1cm} = (-\ii\,\hbar)^3 \, \left(\frac{g}{3!}\right)^2 \ {\widetilde{\sgreen}}(p_1) \ {\widetilde{\sgreen}}(p_2) \ {\widetilde{\sgreen}}(p_3) \ {\widetilde{\sgreen}}(p_4) \ \e^{-\frac{\ii\,\lambda}{2}\, \sum\limits_{a<b}\, (p_{az}\, \ell_{b} - p_{bz}\, \ell_{a})} \\
& \hspace{2cm} \times (2\pi)^3 \ \delta(E_1 + E_2 + E_3 + E_4) \ \delta(p_{1z} + p_{2z} + p_{3z} + p_{4z}) \ \delta_{\ell_{1} + \ell_{2} + \ell_{3} + \ell_{4}, 0}    \\
& \hspace{3cm} \times \int_0^\infty\, \alpha_k \, \d \alpha_k \ \left(\frac{\ttF_{\ell_1, \ell_2, \ell_1 + \ell_2} (\alpha_1, \alpha_2, \alpha_k) \, \ttF_{\ell_3, \ell_4, \ell_3 + \ell_4} (\alpha_3, \alpha_4, \alpha_k)}{(E_1 + E_2)^2 + (p_{1z} + p_{2z})^2 - \alpha^2_k + m^2} \right. \\
&\hspace{6cm} + \frac{\ttF_{\ell_1, \ell_3, \ell_1 + \ell_3} (\alpha_1, \alpha_3, \alpha_k) \, \ttF_{\ell_2, \ell_4, \ell_2 + \ell_4} (\alpha_2, \alpha_4, \alpha_k)}{(E_1 + E_3)^2 + (p_{1z} + p_{3z})^2 - \alpha^2_k + m^2}  \\
&\hspace{6cm} + \left. \frac{\ttF_{\ell_1, \ell_4, \ell_1 + \ell_4} (\alpha_1, \alpha_4, \alpha_k) \, \ttF_{\ell_2, \ell_3, \ell_2 + \ell_3} (\alpha_2, \alpha_3, \alpha_k)}{(E_1 + E_4)^2 + (p_{1z} + p_{4z})^2 - \alpha^2_k + m^2} \right) \ .
\end{split}
\end{align}

Diagrammatically, with a clear analogy to the standard case, we represent this result as
\begin{align}
\begin{split}
\widetilde C_n(p_1,p_2, p_3, p_4)_{\mbox{\tiny 1PI}}^{\swzero}  &= \e^{-\frac{\ii\,\lambda}{2}\, \sum\limits_{a<b}\, (p_{az}\, \ell_{b} - p_{bz}\, \ell_{a})}  \\
& \qquad \times \ \left(\footnotesize \begin{tikzpicture}[scale=0.5, baseline]
    \coordinate (L) at (0,0);
    \coordinate (R) at (2,0);
    \draw[decoration={markings, mark=at position 0.5 with {\arrow{Latex}}}, postaction={decorate}] (-2,2) -- (L);
    \draw[decoration={markings, mark=at position 0.5 with {\arrow{Latex}}}, postaction={decorate}] (-2,-2) -- (L);
    \draw[decoration={markings, mark=at position 0.5 with {\arrow{Latex}}}, postaction={decorate}] (L) -- (R);
    \draw[decoration={markings, mark=at position 0.5 with {\arrow{Latex[reversed]}}}, postaction={decorate}] (R) -- (4,2);
    \draw[decoration={markings, mark=at position 0.5 with {\arrow{Latex[reversed]}}}, postaction={decorate}] (R) -- (4,-2);
    \node at (-2.5,2) {$p_1$};
    \node at (-2.5,-2) {$p_2$};
    \node at (4.5,2) {$p_3$};
    \node at (4.5,-2) {$p_4$};
\end{tikzpicture} \normalsize
+ 
\footnotesize
\begin{tikzpicture}[scale=0.5, baseline]
    \coordinate (T) at (1,1);
    \coordinate (B) at (1,-1);
    \draw[decoration={markings, mark=at position 0.5 with {\arrow{Latex}}}, postaction={decorate}] (-2,2) -- (T);
    \draw[decoration={markings, mark=at position 0.5 with {\arrow{Latex}}}, postaction={decorate}] (4,2) -- (T);
    \draw[decoration={markings, mark=at position 0.5 with {\arrow{Latex}}}, postaction={decorate}] (T) -- (B);
    \draw[decoration={markings, mark=at position 0.5 with {\arrow{Latex[reversed]}}}, postaction={decorate}] (B) -- (-2,-2);
    \draw[decoration={markings, mark=at position 0.5 with {\arrow{Latex[reversed]}}}, postaction={decorate}] (B) -- (4,-2);
    \node at (-2.5,2) {$p_1$};
    \node at (-2.5,-2) {$p_2$};
    \node at (4.5,2) {$p_3$};
    \node at (4.5,-2) {$p_4$};
\end{tikzpicture} \normalsize \right. \\
& \hspace{5cm} \left.
+
\footnotesize
\begin{tikzpicture}[scale=0.5, baseline]
    \coordinate (T) at (-0.5,1);
    \coordinate (B) at (-0.5,-1);
    \coordinate (C) at (1,0);
    \draw[decoration={markings, mark=at position 0.5 with {\arrow{Latex}}}, postaction={decorate}] (-2,2) -- (T);
    \draw[decoration={markings, mark=at position 0.5 with {\arrow{Latex}}}, postaction={decorate}] (-2,-2) -- (B);
    \draw[decoration={markings, mark=at position 0.5 with {\arrow{Latex}}}, postaction={decorate}] (T) -- ($(C)+(-0.15,0.1)$);
    \draw[decoration={markings, mark=at position 0.5 with {\arrow{Latex}}}, postaction={decorate}] ($(C)+(+0.15,-0.1)$) -- (4,-2);
    \draw[decoration={markings, mark=at position 0.5 with {\arrow{Latex}}}, postaction={decorate}] (B) -- (4,2);
    \draw[decoration={markings, mark=at position 0.5 with {\arrow{Latex}}}, postaction={decorate}] (T) -- (B);
    \node at (-2.5,2) {$p_1$};
    \node at (-2.5,-2) {$p_2$};
    \node at (4.5,2) {$p_3$};
    \node at (4.5,-2) {$p_4$};
\end{tikzpicture} \normalsize \right) \ .
\end{split}
\end{align}

\section{Final remarks}
\label{sec:4}

In this contribution we discussed two different quantization schemes in the algebraic BV formalism: the standard (Feynman) quantization and  braided quantization. While the standard quantization can be performed in any basis,  braided quantization naturally selects a basis adapted to the Drinfel'd twist defining the noncommutative deformation. More precisely, in the braided BV formalism all maps entering a homotopy equivalence must be covariant with respect to the deformed Poincar\'e symmetry induced by the twist. In our analysis, we carried out the standard quantization in the plane wave basis (\ref{BazisPl}), whereas the braided quantization uniquely selects the basis of cylindrical harmonics  (\ref{BasisCyl}).

As expected from previous work \cite{AngTwistMi, Wallet1, Wallet2, UsBessel}, the standard quantization leads to two inequivalent classes of diagrams in tree-level four-point functions, with one diagram in each of the $s$-, $t$- and $u$-channels for each class. On the other hand, the braided quantization yields only a single class of diagrams, again with one diagram in each of the $s$-, $t$- and $u$-channels, and the noncommutativity enters only through a phase factor depending on the external momenta.

In \cite{UsBessel} and in this contribution we have considered only the scalar $\phi^3$-theory on $\lambda$-Minkowski space. It would be very interesting to extend this analysis to gauge theories on $\lambda$-Minkowski space and to understand the consequences of the deformed Poincar\'e symmetry as well as the corresponding deformed conservation laws in that setting. Another important open problem is to understand how to systematically extract the one-particle irreducible diagrams from the full set of diagrams (disconnected, reducible and irreducible) that arise from the definition of correlation functions using homological perturbation theory, see for example (\ref{G4def}), and how to construct scattering amplitudes within this formalism based on a homotopy algebraic version of the LSZ reduction theorem in textbook quantum field theory.

\paragraph{Acknowledgments.}

We thank the organisers of the Corfu Summer Institute
2025 for the stimulating meeting and the opportunity to present our work. The work of DjB, MDC and SDj is supported by Project 451-03-34/2026-03/200162 of the Serbian Ministry of Science, Technological Development and Innovation. Support form the COST Actions CaLISTA CA21109 and THEORY-CHALLENGES CA22113 is also acknowledged.

\end{document}